\documentclass{PoS}

\title{David vs. Goliath: pitfalls and prospects in abundance analyses of dwarf vs. giant stars}

\ShortTitle{Abundance analyses of dwarf vs.\ giant stars}

\author{\speaker{Andreas J.\ Korn}\\
        Division of Astronomy and Space Physics, Department of Physics and Astronomy, Uppsala University, Sweden\\
        E-mail: \email{andreas.korn@fysast.uu.se}}

\abstract{I review some of the current limitations in modelling stellar atmospheres of solar-type stars and approaches to overcome them. }

\FullConference{11th Symposium on Nuclei in the Cosmos, NIC XI\\
        July 19-23, 2010\\
        Heidelberg, Germany}

\begin{document}

\section{Introduction}
Impressive abundance results, e.g. for $r$-process-enhanced giant stars (Sneden, these proceedings), were presented at this conference. Quantitative stellar spectroscopy seems to be able to yield hard boundary conditions for nuclear physics. Abundance trends practically without cosmic scatter have been uncovered down to metallicities of [Fe/H]\,=\,$-4$ (see the case of chromium in giants, Cayrel et al.\ 2004). Let us, however, not forget that chemical surface abundances inevitably reflect physical processes taking place in stars. For studies of Galactic chemical evolution, the most relevant processes are those that alter the surface abundances making stars imperfect data carriers.

The stellar spectroscopist infers the chemical composition of stars using models usually containing a great number of simplifying assumptions. This sentence contains the often made simplifying assumption that one can equate the surface composition with that of the star as a whole. For ordinary Population I/II main-sequence stars more or less like the Sun ($T_{\rm eff} = 5800 \pm 500$\,K), this assumption is violated at the 20-60\,\% level (e.g. Korn et al.\ 2007, Mel\'{e}ndez et al.\ 2009). In the following, I shall discuss a few of these assumptions highlighting those leading to sizable effects. What is sizable in this context?

\section{Setting the scale}
On what level do we need to be worried about the validity of derived chemical abundances? The answer to this question will depend very heavily on the application envisioned. It is, however, clear that it is easier to achieve  good relative abundances (precision) than absolute ones (accuracy). Furthermore, a higher level of precision can be safeguarded by analysing homogeneous samples of stars. In this respect, a sample of stars consisting of dwarfs and giants may be problematic.

From a user perspective, it is less interesting from which stars the abundances were inferred. The chemical-evolution modeller tries to reproduce structure(s) in abundance-abundance diagrams. At a given metallicity, such structures are usually present on or below the 0.3\,dex level in logarithmic abundance. Fuhrmann (2004) has shown that thick-disk stars typically have 0.3\,dex less iron at a given magnesium abundance (cf.\ his Fig.~34). Nissen and Schuster (2010) give evidence for two halo populations differing in the abundances of the $\alpha$-elements by less than 0.2\,dex at a given metallicity. If we want to learn something about chemical structures of this sort, then an abundance precision on the same scale may suffice for statistical samples of stars. For a star-by-star classification, a precision 5-10 times higher should be aimed for (Fuhrmann's and Schuster \& Nissen's differential abundances are good to 0.03\,dex).

\section{Dwarf vs.~giants}
There are numerous advantages that dwarfs have over giants, and vice versa. The evolution on the main sequence is slow, this evolutionary phase is thus well-represented in the Hertzsprung-Russell diagram, even of very old populations (but see Bromm, these proceedings, regarding Population III). Number statistics is, however, only one side of the coin. There are typically three magnitudes in visual brightness between a main-sequence turn-off star and a red giant below the bump. At 2\,kpc, such an RGB star shines at 13th magnitude, within spectroscopic reach of 4m-class telescopes. A turn-off star at the same distance is, in practical terms, an 8m application. And in Galactic terms, 2\,kpc does not really get you into the Galactic halo.

Cool giants are not only more luminous, the lower densities in their atmospheres favour the formation of spectral lines in the majority species (like in Fe\,{\sc ii}). It is in these stars that we can behold the full glory of the $r$-process pattern. Rare, even radioactive species like uranium can be found under favourable circumstances (Cayrel et al.\ 2005) allowing nucleo-chronometric age dating. The lower photospheric temperatures favour the formation of molecules which allow us to probe isotopic ratios (e.g.\ C-12/C-13, Mg-25/Mg-24, see e.g.\ Yong et al.\ 2004). But the evolution to higher luminosities comes at a price: dredge-up episodes take place mixing newly fused elements (He, C and N) to the surface. Lithium is burned in this process. Some of the most abundant and interesting elements can thus not be studied in giants, other than to learn about stellar structure and evolution. The uncertainties related to when exactly the dredge-up episodes take place (isotopic ratios are sensitive probes of this) and what the structural consequences are (e.g.\ the effect of the helium abundance on the surface gravity) add to the complexity of interpreting abundance data of giant stars.

Given that dwarfs eventually become giants, should one not expect chemical abundances of, say, iron-peak elements from both groups to agree? The answer is yes and no. For one thing, one has to make sure one samples the same population (not always a given in magnitude-limited surveys). For another, the limitations inherent to our models may affect dwarfs and giant differently. Additional biases may arise from physics that is not included in our modelling. Below, a few such effects beyond classical modelling are briefly discussed.

\section{Non-LTE}
I wrote an invited review on this topic two years ago (Korn 2008). The situation is essentially unchanged: we know that non-LTE plays a significant role in the line formation of minority species like Fe\,{\sc i}, to a lesser extent also for majority species. But this very general statement pays no justice to the complexity of the specific atom. For many elements, non-LTE effects remain unexplored. For all but the lightest species, there are sizable uncertainties in the computations stemming from the unknown strength of collisions with neutral hydrogen. The choices made in connection with hydrogen collisions often decide on the overall strength of departures from LTE.

We have, for lack of a better theory, taken an observational approach, calibrating the strength of hydrogen collisions by means of well-studied stars with significant HIPPARCOS parallaxes. This tends to work very well (see Korn et al.\ (2003) for iron, Mashonkina et al.\ (2007) for calcium). While this approach has been criticized for subsuming all modelling biases into a single, poorly modelled collisional strength, it may, for the time being, be better than ignoring such collisions altogether. Proper quantum-mechanical calculations are underway (e.g. Barklem et al.\ (2010) for Na+H), but it will likely take 5-10 years until a complex atom like iron can be tackled.

Until disproven, it cannot be ruled out that mismatches between dwarfs and giants (see e.g.\ Bonifacio et al.\ 2009 for the case of chromium) are due to our simplistic modelling in LTE. Until this is cleared up, modelling such LTE-based abundance trends in terms of Galactic chemical evolution may be futile.

\section{Hydrodynamics}
The realization that full account of hydrodynamics significantly changes the $T-\tau$-relations of solar-type stars has led to rather drastic changes in stellar abundances, not the least for metal-poor stars. At [Fe/H]\,=\,$-3$, adiabatic cooling lowers the temperatures in the upper photospheres by a couple of thousand K, with consequences for all lines formed in these layers (neutral species, molecules). Effects can be large in both dwarfs and giants and reach up to 1\,dex for molecules. Asplund (2005) gives a competent overview of the subject.

Coupling of hydrodynamics and non-LTE is nowadays possible, at least for atoms of moderate complexity. Curiously, 1D--LTE and 3D--non-LTE of lithium practically coincide for stars on the Spite plateau (Barklem et al.\ 2003). This does, however, likely not hold when analysing cooler stars. So trends of lithium with effective temperature may well be affected by 3D--non-LTE. This should be explored further, as it may tell us how well we understand the structural changes in connection with the first dredge-up.

\section{Atomic diffusion}
It has been speculated for decades that the surface abundances of metal-poor halo dwarfs are systematically affected by gravitational settling and radiative levitation (e.g.\ Michaud, Fontaine \& Baudet 1984), collectively referred to as atomic diffusion. In this physical picture, helium and lithium would settle appreciably (with important effects for stellar ages), while elements like chlorine or potassium could be accumulated in the atmosphere to abundance levels above the composition of the gas from which the star once formed. These effects are predicted to be prominent whenever convection is weak. It is thus not surprising that metal-poor turn-off stars of spectral type F would be significantly affected (also from the point of view of their proximity to all the classes of chemically peculiar stars between early F and late B).

More than 20 years of research into atomic diffusion in solar-type stars made it clear that uninhibited diffusion is unlikely to occur: its abundance effects would have been seen, and the Spite plateau of lithium would be distorted at the hot end. Extra mixing below the convective envelope was introduced as a moderating process, albeit without specifying the physical mechanism (Proffitt \& Michaud 1991).

A careful differential analysis of unevolved and evolved stars in the metal-poor globular cluster NGC 6397 ([Fe/H]\,=\,$-2.1$) revealed systematic trends of abundance with evolutionary stage (Korn et al.\ 2006, 2007). The abundance trends are not overwhelmingly large (up to 0.2\,dex), but there is excellent agreement of the element-specific amplitude of the trends with predictions of stellar-structure models with atomic diffusion. The efficiency of the postulated extra mixing is found to be close to the lowest efficiency required to keep the Spite plateau thin and flat (cf.\ Richard et al.\ 2005). The inferred initial lithium abundance is $\log \varepsilon$(Li)\,=\,2.54\,$\pm$\,0.1, in good agreement with WMAP-calibrated BBN predictions (cf.\ Steigman, these proceedings).

Mel\'{e}ndez et al. (2010) took a fresh look at the Spite plateau among field stars comparing its morphology to atomic-diffusion predictions on a stellar-mass scale (rather than on the customary metallicity scale). They confirm that the Spite plateau is shaped by atomic diffusion and that stellar physics alone can account for the cosmological lithium-7 problem.

We (Nordlander, Korn \& Richard) are currently investigating whether or not there is a (hypothetical) effective-temperature scale which fully explains the cosmological lithium-7 problem, is compatible with the white-dwarf cooling age for NGC 6397 (11.5\,$\pm$\,0.5\,Gyr, Hansen et al.\ 2007) and describes the abundance trends at a high mixing efficiency \`{a} la Mel\'{e}ndez et al. (2010). Seemingly, such a scale does not exist. More work is needed here.

\begin{figure}
\includegraphics[width=0.9\textwidth,clip=]{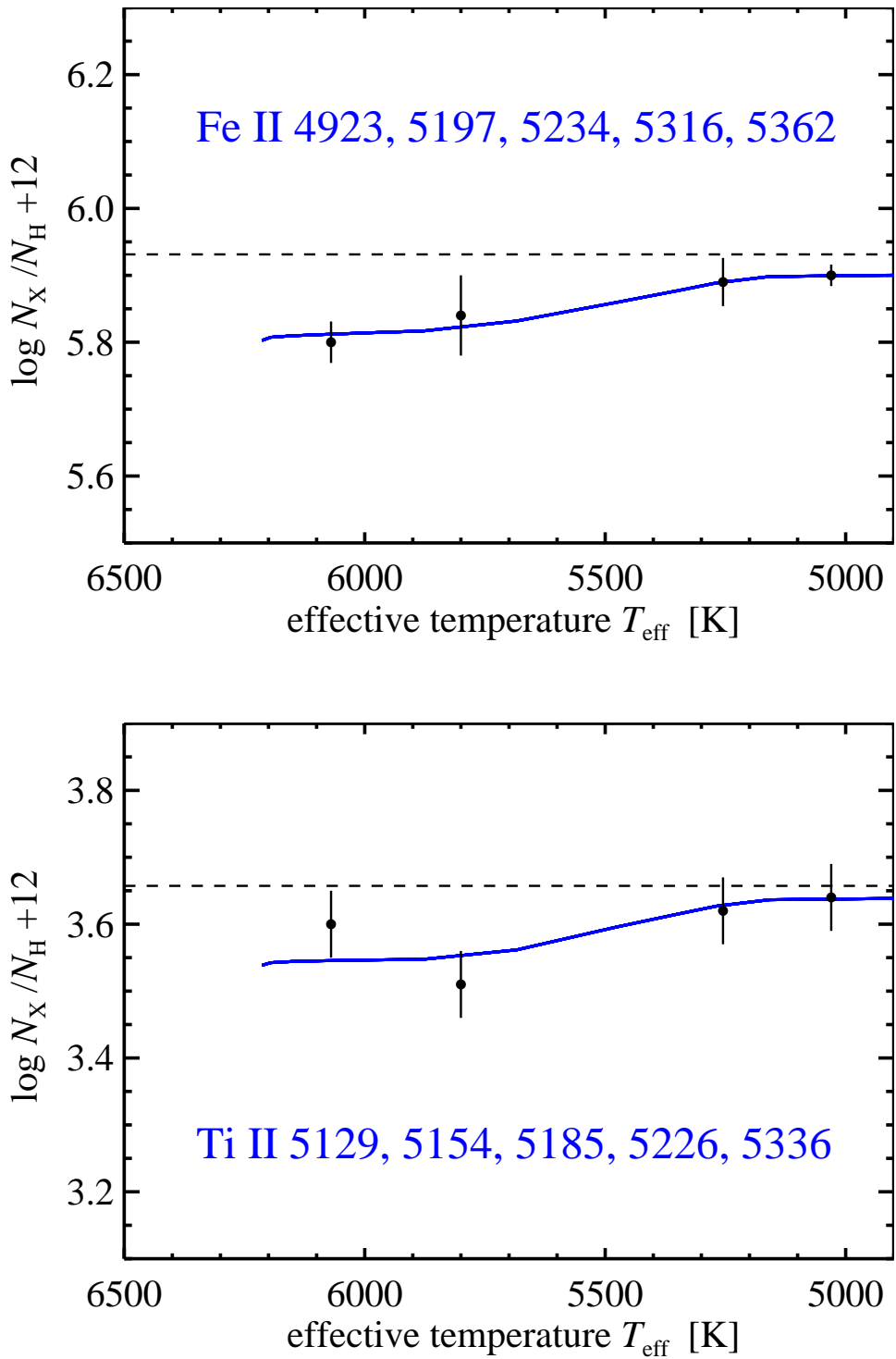}
\caption{1D--LTE trends of iron and titanium in the metal-poor globular cluster NGC 6752, from stars near the turn-off to the RGB (bullets), compared to stellar-structure predictions with atomic diffusion (lines). Stellar parameters are photometric, with surface-gravity differences tightly constrained by the luminosity differences of the stars. Abundances are based on the lines listed. The lines originate from the log\,$g$-sensitive majority species for which both non-LTE and 3D corrections are expected to be small.}
\end{figure}

Clusters at both lower and higher metallicities are being scrutinized for atomic-diffusion signatures. This is observationally challenging, as low-reddening metal-poor globular clusters tend to be further away than NGC 6397. In M 92, turn-off stars have $V\,\approx$\,18.3. One really needs a 10m telescope like Keck to take decent spectra of such stars.

Figure 1 shows a preliminary analysis of stars in NGC 6752 at [Fe/H]\,=\,$-1.6$ taken with VLT/FLAMES-UVES. Systematic trends seem to exist between the groups of stars, but the overall amplitude is low, reaching 0.1\,dex for iron. The trends are thus compatible with predictions from stellar-structure models including atomic diffusion, radiative acceleration and extra mixing below the convective envelope, with rather efficient extra mixing.

It is interesting to speculate what happens at lower metallicities. If the two data points constraining the turbulent-mixing efficiency (NGC 6752 @ $-1.6$ and NGC 6397 @ $-2.1$) constitute a trend, then one would expect relatively inefficient extra mixing in the most metal-poor globular clusters. This would lead to sizable abundance differences between dwarfs and giants, in excess of 0.3\,dex for certain elements like magnesium. This is currently being tested at the VLT (M 30, Lind et al.) and at Keck (M 92, Cohen et al.). At even lower metallicities, inefficient extra mixing could potentially explain the breakdown of the Spite plateau (Sbordone et al.\ 2010), as only models within a certain range of extra-mixing efficiencies can produce a thin and flat Spite plateau (see Richard et al.\ 2005).

\section{{\sl \textbf{Tertium non datur?}}}
Indeed, there is more to stellar astrophysics than dwarfs and giants. In particular, subgiants deserve to be mentioned as a third and intermediate group of stars, as they seem to combine some of the best properties of both the less evolved and the more evolved objects: more luminous than dwarfs, but still rather high-gravity objects, not yet affected by the first dredge-up. The main showstopper has so far been the fact that they are relatively rare. But this will no longer be an obstacle in the era of all-sky surveys like Gaia.

Subgiants really come into their own right when we talk stellar ages. A beautiful example is given by Bernkopf, Fiedler \& Fuhrmann (2001): the few known thick-disk subgiants with significant HIPPARCOS parallaxes seem to be systematically older than comparable thin-disk stars, indicating a possible hiatus in star formation between thick- and thin-disk formation of several Gyr.

In the end, the ambition has to be to reliably derive the chemical compositions of dwarfs and giants alike, across the full range of metallicities, with full account of hydrodynamics, non-equilibrium line formation, atomic diffusion and dredge-up. The ability to combine kinematical, chemical and age information for carefully selected subsets of stars ({\sl tertium datur}: subgiants!) will undoubtedly propel us into the era of Precision Galactic Archaeology.

\acknowledgments
I would like to thank Frank Grundahl and Olivier Richard who both provided crucial input data used in the preliminary analysis of dwarf-to-giant stars in NGC 6752.

\end{document}